\documentclass{iopart}
\usepackage{epsfig}
\usepackage{subfigure}
\begin{document}
 
\article[Solution of the RHIC HBT puzzle]
  {Strangeness in Quark Matter 2008, Beijing, China}
  {Solution of the RHIC HBT puzzle with Gaussian initial conditions \footnote{Supported in part by the Polish Ministry of Science and Higher Education, grants N202 153 32/4247 and N202 034 32/0918, and by the U.S. NSF Grant No. PHY-0653432.}
}  
 
\author{W. Florkowski$^{1,2}$, W. Broniowski$^{1,2}$, M. Chojnacki$^1$, A.~Kisiel$^{3,4}$ }

\address{$^1$ The H.~Niewodnicza\'nski Institute of Nuclear Physics, 
Polish Academy of Sciences, 31-342 Krak\'ow, Poland} 
\address{$^2$ Institute of Physics, Jan~Kochanowski University, 25-406 Kielce, Poland}
\address{$^3$ Faculty of Physics, Warsaw University of Technology, 00-661 Warsaw, Poland}
\address{$^4$ Department of Physics, Ohio State University, Columbus, OH 43210, USA}
   
\begin{abstract}
It is argued that the consistent description of the transverse-momentum spectra, elliptic flow, and the HBT radii in the relativistic heavy-ion collisions studied at RHIC may be obtained within the hydrodynamic model if one uses the Gaussian profile for the initial energy density in the transverse plane. Moreover, we show that the results obtained in the scenario with an early start of hydrodynamics (at the proper time $\tau_0$ = 0.25 fm) are practically equivalent to the results obtained in the model where the hydrodynamics is preceded by the free-streaming stage of partons (in the proper time interval $0.25 \,\hbox{fm} \leq \tau \leq 1 \, \hbox{fm}$) which suddenly equilibrate and with the help of the Landau matching conditions are transformed into the hydrodynamic regime (at the proper time $\tau_0$ = 1 fm).
\end{abstract} 
%\submitto{\JPG}
%\maketitle 

\bigskip

The problem of the consistent description of various features of the soft hadron production in the nucleus-nucleus collisions at the Relativistic Heavy-Ion Collider (RHIC) is well known \cite{Heinz:2002un}. In particular, the so called RHIC HBT puzzle \cite{Heinz:2002un,Hirano:2004ta,Lisa:2005dd,Huovinen:2006jp} refers to the difficulty of simultaneous description of the hadronic transverse-momentum spectra, the elliptic flow coefficient $v_2$, and the Hanbury-Brown--Twiss (HBT) interferometry data in various approaches including hydrodynamics \cite{Heinz:2001xi,Hirano:2001yi,Hirano:2002hv,Zschiesche:2001dx,Socolowski:2004hw}. 

The observable which is particularly difficult to reproduce in the hydrodynamic calculations is the ratio of the out to side radius for the pions, $R_{\rm out}/R_{\rm side}$. In the hydrodynamic approach this ratio reaches typically the values around 1.5, while the experimental values approach unity (for the average transverse-momentum of the pion pair $k_T \sim$ 0.5 GeV). The experimental results are usually interpreted as the indication for the strong transverse flow and the relatively short emission time. This in turn suggests that the equation of state cannot be too soft. In fact, the use of the semi-hard equation of state constructed in Ref. \cite{Chojnacki:2007jc}, which is based on the smooth interpolation between the hadron-gas model and the lattice data, helps to reduce the model values of the ratio $R_{\rm out}/R_{\rm side}$ to 1.20-1.25 \cite{Chojnacki:2007rq}. At the same time, the transverse-momentum spectra and $v_2$ of pions, kaons, and protons are well reproduced. 

Besides the use of the appropriate equation of state, with the temperature dependent sound velocity which remains quite substantial at the vicinity of the critical temperature $T \sim T_c$, also the use of the complete set of hadronic resonances turns out to be responsible for  the good description of the data achieved in Ref. \cite{Chojnacki:2007rq}. The inclusion of the feeding from all known hadronic resonances to the yields of light particles (such as pions, kaons and protons) is an important feature responsible for the success of the thermal models of heavy-ion collisions. It also allows for the unified description of the chemical and kinetic freeze-out \cite{Broniowski:2001we}. The single freeze-out assumption shortens the emission time and also helps to reproduce $R_{\rm out}/R_{\rm side}$. We note that in our approach the hadron emission at freeze-out is modeled  with the Monte Carlo code {\tt THERMINATOR} \cite{Kisiel:2005hn}, which implements the two-particle method of the calculation of the correlation function with or without the Coulomb corrections. 

\begin{figure}[tb]
\begin{center}
\includegraphics[angle=0,width=0.7 \textwidth]{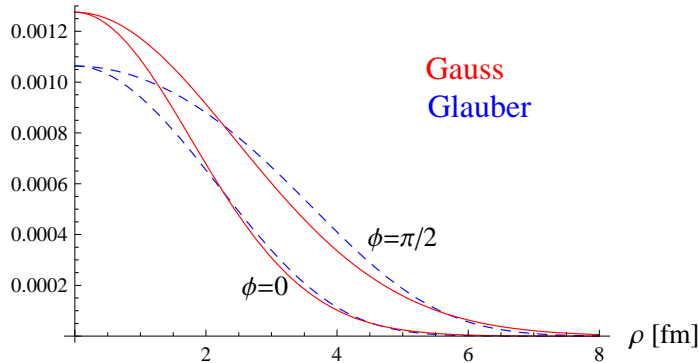}
\end{center}
\vspace{-6.5mm}
\caption{In-plane and out-of-plane sections of the two-dimensional  energy-density profiles for the centrality class $c=30-40\%$ obtained from the Glauber model (dashed lines) and the Gaussian  parametrization used in Ref. \cite{Broniowski:2008vp}. Both profiles are normalized to unity,  and in addition
$\langle x^2 \rangle=a^2$, $\langle y^2 \rangle=b^2$.  
\label{fig:shape}}
\end{figure}

In the recent work \cite{Broniowski:2008vp}, we have shown that the further improvement of the description of the data may be achieved if the standard initial conditions, following directly from the optical Glauber model, are replaced by the Gaussian energy-density profiles of the form
\begin{eqnarray}
n(x,y)=\exp \left ( -\frac{x^2}{2a^2} -\frac{y^2}{2 b^2} \right ).
\label{profile}
\end{eqnarray} 
It is important to emphasize that the Gaussian profiles do originate from the Glauber model -- they are Gaussian fits to the source distribution determined by the Monte Carlo version of the Glauber model \cite{Broniowski:2007nz}. 

\begin{figure}[t]
\begin{center}
\includegraphics[angle=0,width=1.0 \textwidth]{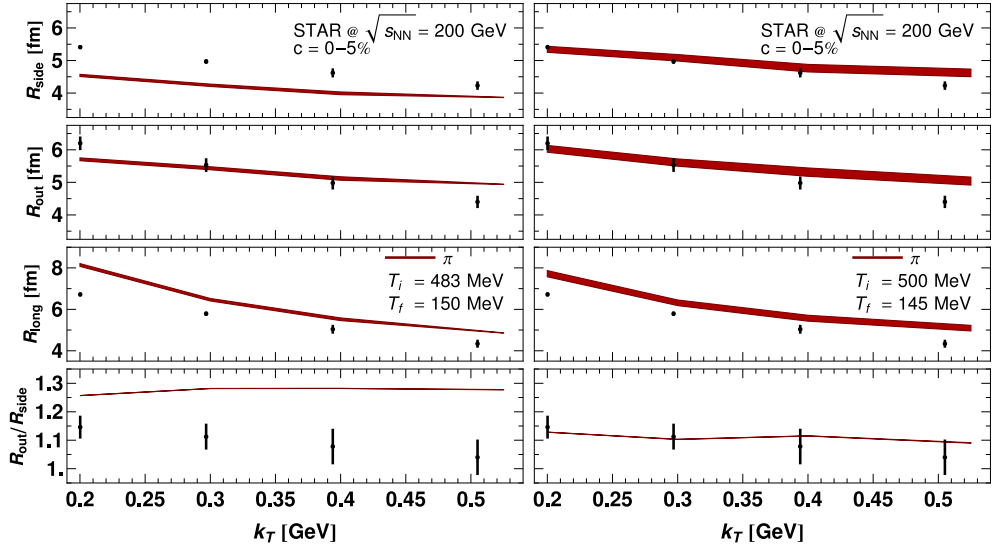}
\end{center}
%\vspace{-6.5mm}
\caption{ The pion HBT radii $R_{\rm side}$ , $R_{\rm out}$ , $R_{\rm long}$,  and the ratio $R_{\rm out}/R_{\rm side}$ for central collisions, shown as the functions of the average momentum of the pair and compared to the RHIC Au+Au data. The left panel illustrates our best results obtained with the standard Glauber initial conditions \cite{Chojnacki:2007rq}, while the right panel illustrates the results obtained with the Gaussian initial conditions \cite{Broniowski:2008vp}. The data from \cite{Adams:2004yc}.
\label{fig:hbt}}
\end{figure}

The examples of the in-plane and out-of-plane profiles corresponding to the optical Glauber model and the Gaussian parameterization (\ref{profile}) are presented in Fig. \ref{fig:shape}. We note that both types of profiles yield the same values of $\langle x^2 \rangle = a^2$ and $\langle y^2 \rangle = b^2$. The main difference between the two profiles is that the Gaussian profile is steeper in the central part. This feature causes the faster development of the transverse flow. Since the magnitude of the transverse flow at freeze-out is constrained  by the experimental data, the faster development of the flow means that the evolution is faster and the duration of the emission process is shorter.  This is once again the requested feature which helps to reproduce correctly the ratio $R_{\rm out}/R_{\rm side}$ -- see Fig. \ref{fig:hbt}. 

\begin{figure}[t]
\vspace{2mm}
\begin{center}
\includegraphics[angle=0,width=0.6 \textwidth]{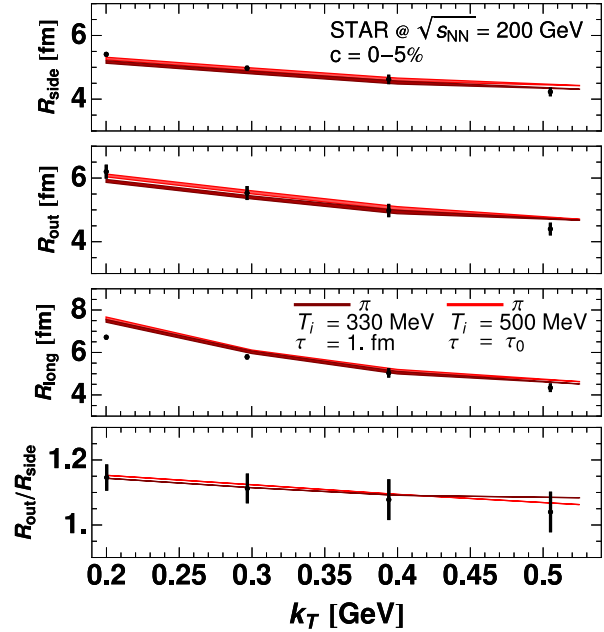}
\end{center}
\vspace{-1mm}
\caption{(Color online) The pion HBT radii $R_{\rm side}$ , $R_{\rm out}$ , $R_{\rm long}$, and the ratio $R_{\rm out}/R_{\rm side}$ for central collisions. The darker (lighter) lines describe the results with (without) FS+SE. The data from \cite{Adams:2004yc}.
\label{fig:hbt2}}
\end{figure}

Our results presented in Fig. \ref{fig:hbt} were obtained  with the assumption that the hydrodynamic evolution starts at the proper time $\tau_0$ = 0.25 fm. In the very recent work \cite{Broniowski:2008qk}, we have shown that the start of the hydrodynamic evolution may be postponed to $\tau$ = 1 fm if the hydrodynamics is preceded by the free-streaming stage of partons (in the proper time interval $0.25 \,\hbox{fm} \leq \tau \leq 1 \,\hbox{fm}$) which suddenly equilibrate at \mbox{$\tau$ = 1 fm}. In such FS+SE scenario (free-streaming + sudden equilibration, see \cite{Kolb:2000sd,Gyulassy:2007zz}), the very fast but delayed equilibration is described with the help of the Landau matching conditions for the energy-momentum tensor. Figure \ref{fig:hbt2} illustrates that the results obtained in the case when the  hydrodynamic evolution starts at $\tau_0$ = 0.25 fm are practically indistinguishable from the results obtained within the FS+SE scenario. This observation suggests that the consistent description of the soft physics at RHIC might be achieved in the approach where the thermalization and the formation of the transverse flow happens gradually within 1 fm. 

In conclusion we state that the consistent and uniform description of the soft hadronic data at RHIC (including the transverse-momentum spectra, the elliptic flow coefficient $v_2$, the azimuthally sensitive HBT, and non-identical particle correlations \cite{Broniowski:2008vp,Kisiel:2008ws,Kisiel:WPCF}) may be achieved within the hydrodynamic approach if a proper choice of the initial profile and a realistic equation of state are used. Other aspects such as the fluctuations of the initial eccentricity \cite{Socolowski:2004hw,Voloshin:2008dg}, inclusion of the resonances, and the two-particle method used for the calculation of the correlation function play also an important role. Moreover, the start of the hydrodynamic evolution may be delayed if the hydrodynamic phase is preceded with the free-streaming stage followed by the sudden equilibration. 

We note that a similar conclusion has been also reached recently in Ref. \cite{Pratt:2008qv} (see also \cite{Lisa:2008gf}) where, however, the fast building of the transverse flow that is required for the correct description of the HBT radii is achieved not by the modification of the initial conditions but by the very early start of the hydrodynamics ($\tau_0$ = 0.1 fm) and the inclusion of the viscous effects. 

\bigskip

%\bibliography{ref-rr}

\end{document}